\documentclass{mem}
\usepackage{natbib}\usepackage{txfonts}\usepackage{balance}
\usepackage{graphicx}
\usepackage[a4paper,breaklinks,dvipdfm]{hyperref}
\idline{83}{0}
\def \th {\thinspace}

\def\approxgt{\mathrel{\hbox{\rlap{\lower.55ex \hbox {$\sim$}} \kern-.3em \raise.4ex \hbox{$>$}}}}
\def\lesssim{\mathrel{\hbox{\rlap{\lower.55ex \hbox {$\sim$}} \kern-.3em \raise.4ex \hbox{$<$}}}}
\def\approxlt{\mathrel{\hbox{\rlap{\lower.55ex \hbox {$\sim$}} \kern-.3em \raise.4ex \hbox{$<$}}}}

\begin{document}

\title{
A review of the Z-track sources
}
\subtitle{}
\author{
M. J. \, Church\inst{} 
\and M. \, Ba\l uci\'nska-Church\inst{}
}
\offprints{M. J. Church}

\institute{
School of Physics \& Astronomy, 
University of Birmingham, Birmingham B15 2TT, UK.
\email{mjc@star.sr.bham.ac.uk}
}

\authorrunning{Church}
\titlerunning{The Z-track sources}

\abstract{The brightest class of low mass X-ray binary source: the Z-track sources are reviewed
specifically with regard to the nature of the three distinct states of the sources.
A physical model is presented for the Cygnus\th X-2 sub-group in which increasing mass accretion
rate takes place on the Normal Branch resulting in high neutron star temperature and radiation
pressure responsible for inner disk disruption and launching of jets. The Flaring Branch
consists of unstable nuclear burning on the neutron star. It is shown that the Sco\th X-1 like 
sub-group is dominated by almost non-stop flaring consisting of both unstable burning and increase
of $\dot M$, causing higher neutron star temperatures. Finally, results of Atoll source surveys 
are presented and a model for the nature of the Banana and Island states in these sources is 
proposed. Motion along the Banana state is caused by variation of $\dot M$. Measurements of 
the high energy cut-off of the Comptonized emission $E_{\rm CO}$ provide the electron 
temperature $T_{\rm e}$ of the Comptonizing ADC; above a luminosity of 
$2\times 10^{37}$ erg s$^{-1}$ $E_{\rm CO}$ is a few keV and $T_{\rm e}$ equals the neutron star 
temperature. At lower luminosities, the cut-off energy rises towards 100 keV showing
heating of the corona by an unknown process. This spectral hardening is the cause of the Island
state of Atoll sources. The models for Z-track and Atoll sources thus constitute a unified model
for low mass X-ray binary sources.
}
\maketitle{}

\section{Introduction}

\begin{figure}[ht!]                                              
\begin{center}
\includegraphics[width=56mm,height=56mm,angle=0]{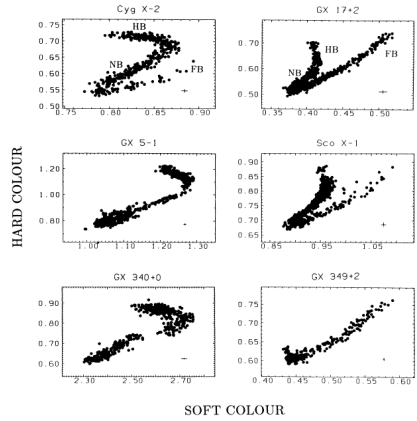}
\caption{\footnotesize
The main Z-track sources: from Hasinger \& van der Klis (1989).}
\label{}
\end{center}
\end{figure}
The characteristic behaviour of the Z-track sources as the brightest group of low mass 
X-ray binaries (LMXB), was demonstrated by Hasinger \& van der Klis (1989) as shown in Fig. 1.
The sources, having luminosities at and above the Eddington limit, display strong physical 
changes via three distinct branches in hardness-intensity or colour-colour diagrams: the 
Horizontal (HB), Normal (NB) and Flaring (FB) branches. There are two sub-groups: the Cygnus\th X-2 
like sources with all three tracks and the Sco\th X-1 like sources in which the HB is weak
but flaring is much stronger. The physical changes have not been understood but the sources 
are important because of the detection of relativistic jets essentially around the hard apex 
between NB and HB. This allows investigation of jet launching by X-ray spectral
studies of conditions near the neutron star when jets are formed. A complete understanding of 
LMXB also requires understanding of the Atoll sources having luminosities less than 
$10^{38}$ erg s$^{-1}$ and displaying essentially two states: the Banana and Island states.

It was proposed by Priedhorsky et al. (1986) and by Hasinger et al. (1990) that the states 
of the Z-track sources might be explained by a monotonic increase of mass accretion rate 
$\dot M$ in the direction HB to NB to FB. Although this was suggested largely by rather
limited evidence for an apparent increase of UV on the FB, the approach has been adopted as 
a standard approach. However, a rather different approach has resulted from the use of a 
particular spectral model based specifically on the extended nature of the Accretion Disk 
Corona (ADC) (Church et al. 2006).
\begin{figure}[h!]                                                   
\begin{center}
\includegraphics[width=56mm,height=46mm,angle=0]{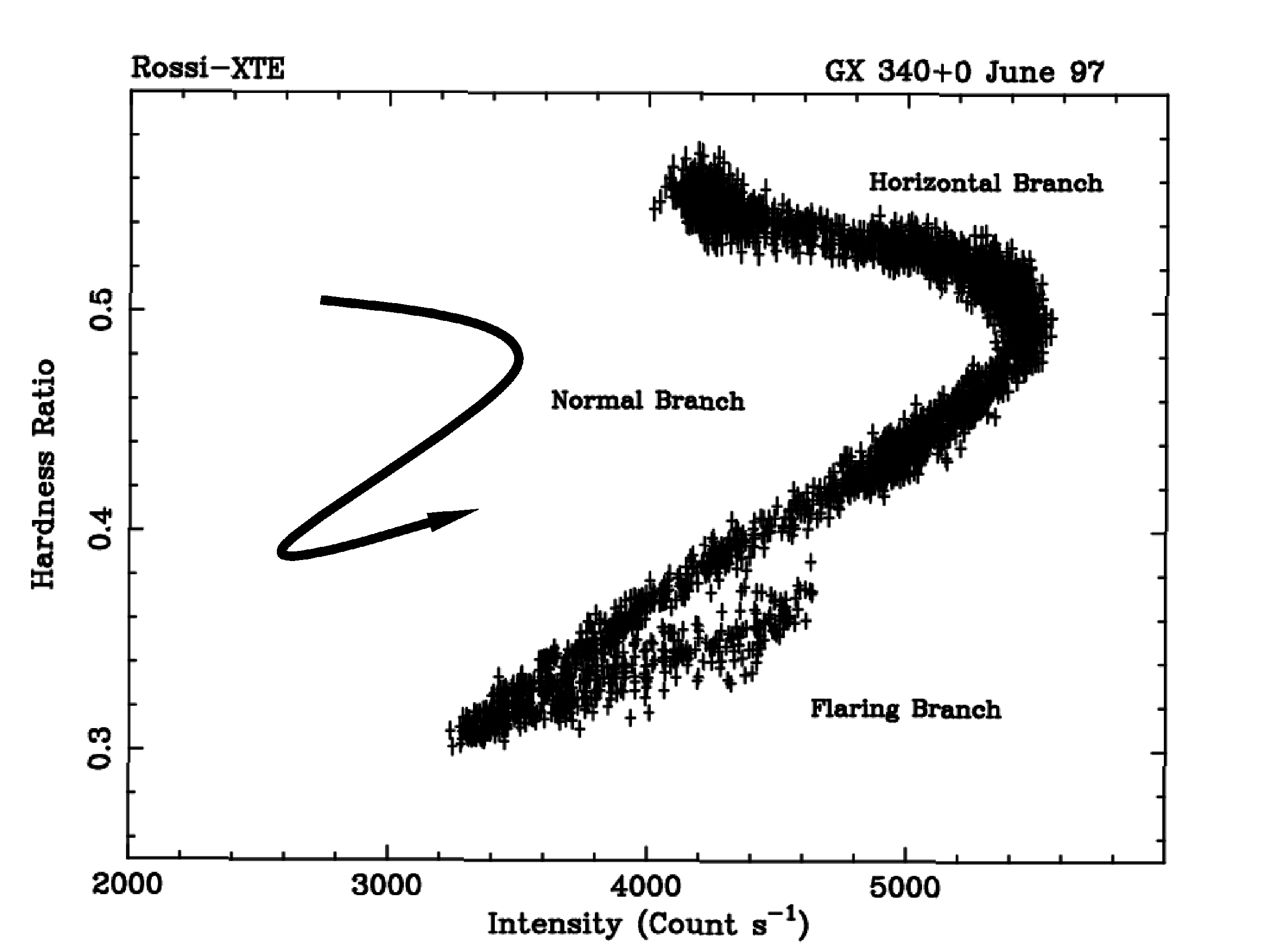}
\caption{\footnotesize A monotonic increase of mass accretion rate.} 
\label{}                                                             
\end{center}                                                         
\end{figure}
\begin{figure}[h!]                                                        
\begin{center}
\includegraphics[width=56mm,height=56mm,angle=270]{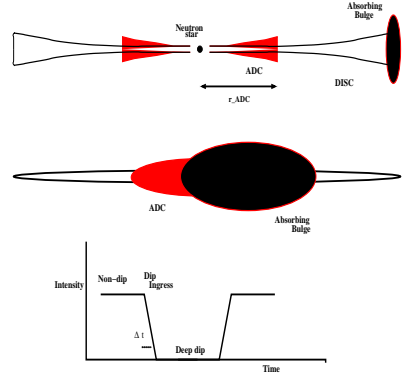}  
\caption{\footnotesize Dip ingress timing: the ingress time allows the size of the extended
Comptonized ADC emitter to be determined.}
\label{}
\end{center}
\end{figure}

The adoption of a realistic model of the X-ray emission is crucial to understanding
and this depends on the evidence for the extended nature of the ADC. Two independent 
techniques have shown that the Comptonizing region in LMXB in general is extended: dip 
ingress timing and also measurement of the Doppler widths of highly ionized emission lines. 
In the high-inclination dipping LMXB, the extended nature of the dominant Comptonized 
emission is proven by its only gradual removal when overlapped by the absorbing bulge 
in the outer disk (Church et al. 1997), and the rate of removal provides the radial extent 
of the ADC of $\sim$ 20\th 000 to 700\th 000 km (Fig. 3; Church \& Ba\l uci\'nska-Church 2004). 
High quality {\it Chandra} grating studies of highly ionized lines in Cyg\th X-2 such as
Fe XXVI, Fe XXV, S XVI and S XIV by Schulz and co-workers (Fig. 4) gave Doppler widths equivalent 
to radial positions of 18\th 000 to 110\th 000, also indicating a very extended ADC 
(Schulz et al. 2009) in excellent agreement with dip ingress timing.
\begin{figure}[h!]                                                       
\begin{center}
\includegraphics[width=66mm,height=66mm,angle=0]{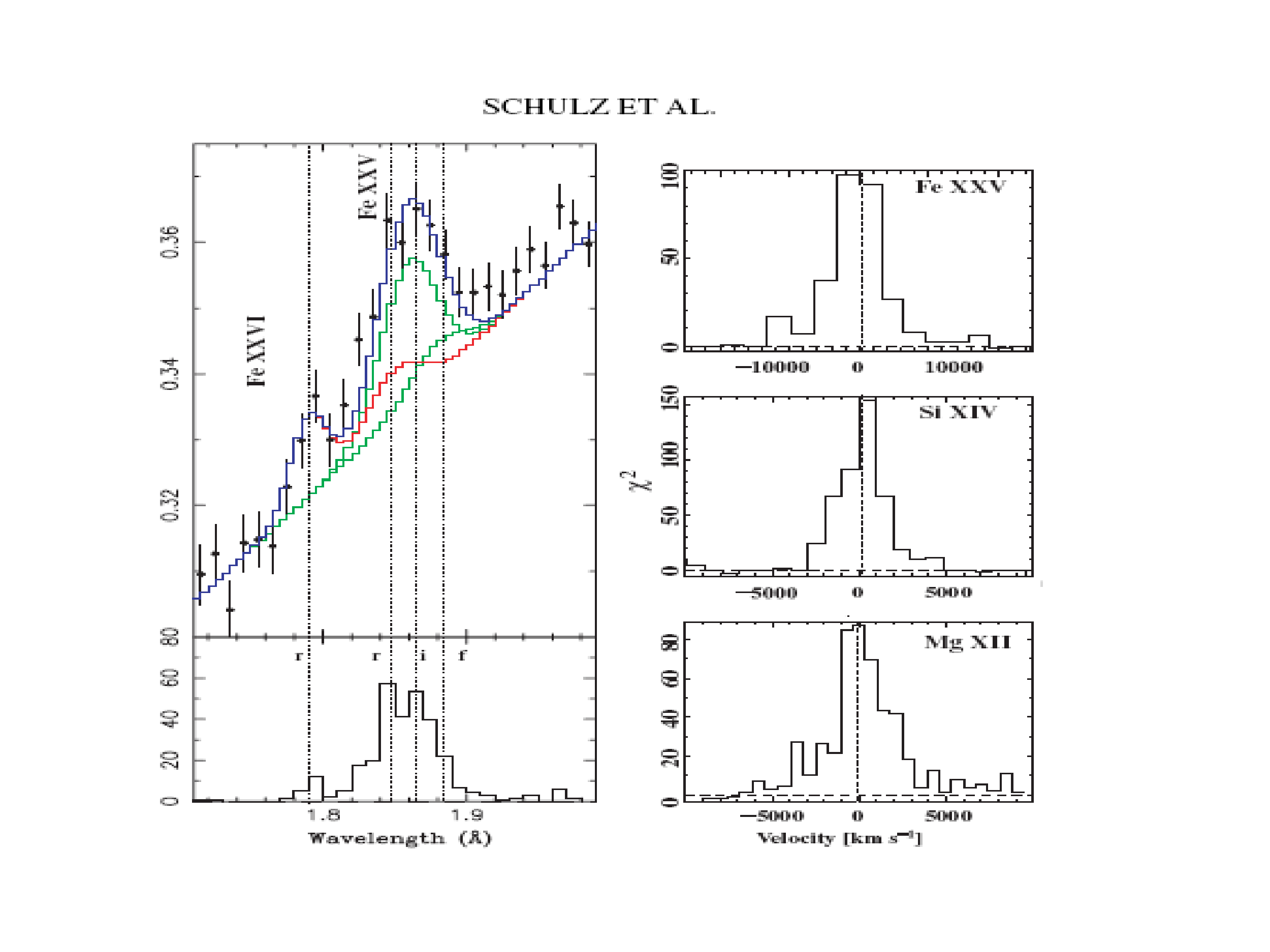}                  
\caption{\footnotesize From Schulz et al. 2009: Doppler widths of highly ionized lines.}
\label{}
\end{center}
\end{figure}

\subsection{Claims that ADC not extended:}

There was originally a view that the site of the dominant Comptonized emission in LMXB
should be a compact central region close to the compact object. As evidence for the extended
ADC emerged, possible ways of avoiding the extended nature were suggested. Highly ionized absorption
features were found in the dipping source X\th 1624-490 by Parmar et al. (2002) and similar features
found in other dipping sources. However, Boirin et al. (2005) and D\'iaz Trigo et al. (2006)
proposed that the X-ray absorption
dips occurring at the orbital period could be modelled by a varying ionization state of absorber,
and that the dipping sources could be explained 
without the need for an extended ADC. However, this did not take into account i) the gradual removal 
of Comptonized emission in dipping which proves an extended nature and ii) dipping would have to be unconnected
with the bulge in the outer disk, contrary to extensive evidence. Thus such claims are not very
physical.

\section{The Cygnus\th X-2 like Z-track sources}

\begin{figure}[h!]                                                    
\begin{center}
\includegraphics[width=56mm,height=56mm,angle=270]{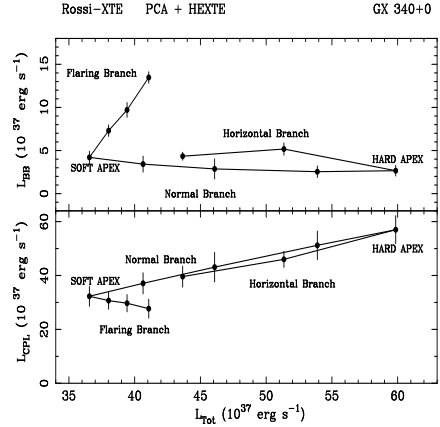}
\caption{\footnotesize The variation of emission component luminosities (Church et al. 2006).}
\label{}
\end{center}
\end{figure}
The spectral model based on an extended ADC has a quite specific form (Church \& Ba\l uci\'nska-Church 2004)
since the seed photons will predominantly arise in the disk beneath the ADC; in addition there
will be thermal emission of the neutron star. It was found that when this model was applied 
in studies of spectral evolution along the Z-track in the Cyg-like sub-group (Cyg\th X-2, 
GX\th 5-1 and GX\th 340+0) the approach provided a rather straightforward explanation (Ba\l uci\'nska-Church 
et al. 2010). The results indicated that evolution along the NB was driven by an increase of $\dot M$
followed by a similar decrease on the HB, while flaring consists of unstable nuclear burning.
\begin{figure}[h!]                                                        
\begin{center}
\includegraphics[width=56mm,height=56mm,angle=270]{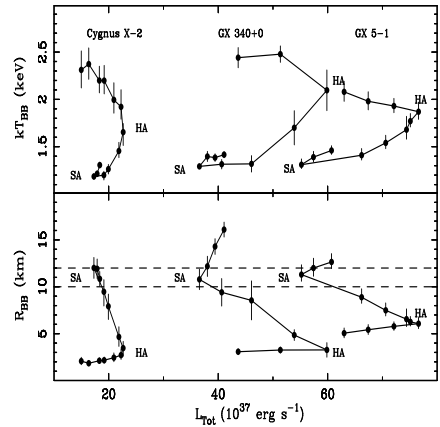}
\caption{\footnotesize Variation of the neutron star blackbody temperature along the Z-track.}
\label{}
\end{center}
\end{figure}
\begin{figure}[h!]                                                        
\begin{center}
\includegraphics[width=56mm,height=26mm,angle=0]{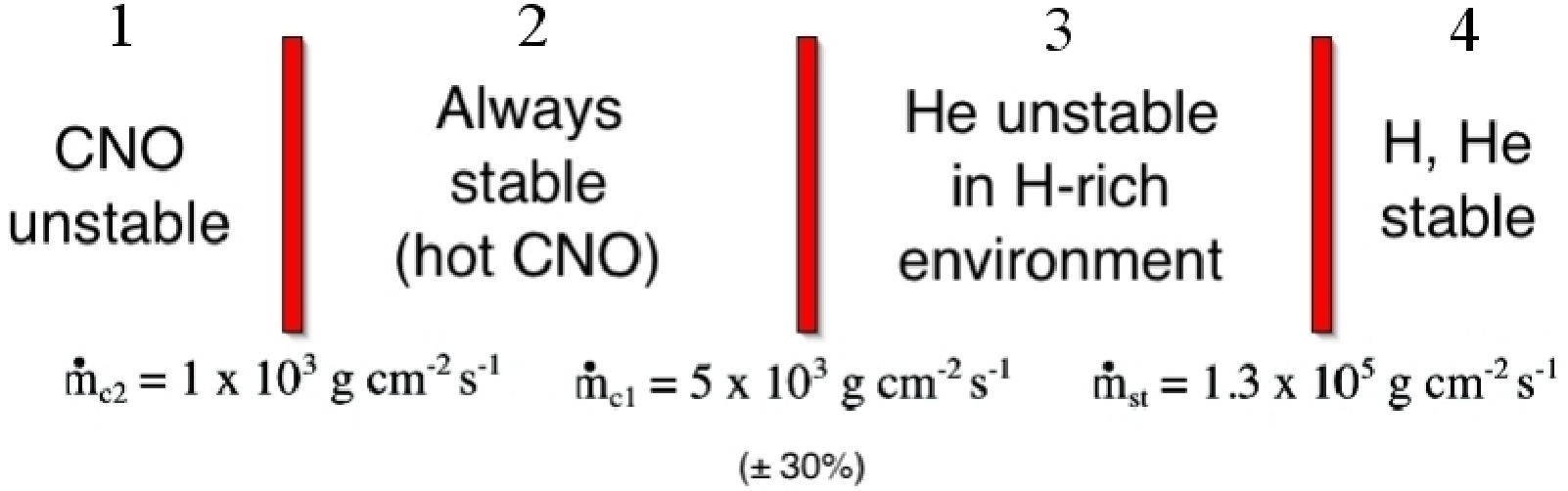}
\includegraphics[width=26mm,height=56mm,angle=270]{church_2011_01_fig07b}       
\caption{\footnotesize Theoretical r\'egimes of stable and unstable nuclear burning, agreeing with
measured values of $\dot m$ (see text) at the onset of flaring.}
\label{}
\end{center}
\end{figure}
A significant result was that spectral fitting showed that 75\% of the X-ray luminosity  is due 
to Comptonized emission as seen in Fig. 5 for GX\th 340+0. Thus the large increase in this ($L_{\rm ADC}$) 
on the NB from the Soft Apex to the Hard Apex strongly suggests that $\dot M$ increases (contrary 
to the $\dot M$ decrease in the standard view). This is supported by the corresponding increase 
of neutron star temperature $kT$ from $\sim$ 1 to more than 2 keV (Fig. 6) which shows that 
the radiation pressure of the neutron star increases by almost 10 times at the Hard Apex. The 
measured neutron star flux becomes several times the Eddington flux and we proposed (Church et al. 2006) 
that this disrupts the inner disk and launches jets at this part of the Z-track by radiation pressure.

The lack of change of $L_{\rm ADC}$ in flaring but the increase of $L_{\rm BB}$ supports the idea that 
$\dot M$ is constant in flaring but that the neutron star luminosity increases due to energy release 
on the surface of the neutrons star. It was found (Ba\l uci\'nska-Church et al. 2010) that the onset of 
flaring coincides with the critical value of $\dot m$ ($\dot M$ per unit area) on the stellar surface for 
unstable nuclear burning (Fig. 7; Bildsten 1998). A schematic view of the physical changes taking place along 
the Z-track is shown in Fig. 8.
\begin{figure}[h!]                                                        
\begin{center}
\includegraphics[width=46mm,height=56mm,angle=270]{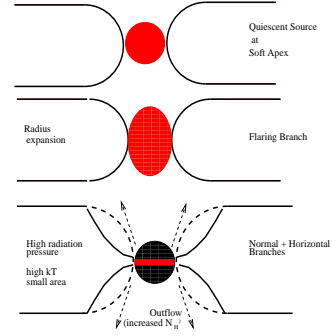}
\caption{\footnotesize Schematic view of the model of the Cygnus\th X-2 like Z-track sources from
Church et al. (2006) and Ba\l uci\'nska-Church et al. (2010).}
\label{}
\end{center}
\end{figure}

\section{Relativistic lines/reflection from the inner disk ?}

A controversial question relevant to these sources, especially Cygnus\th X-2, relates
to whether such features are seen. Done et al. (2002) proposed that Fe emission and reflection
was seen in Cyg\th X-2. More recently, various authors have claimed the detection of
relativistically broadened lines in $\sim$10 LMXB (Cackett et al. 2010 (including Cyg\th X-2); 
Reis et al. 2009; Di Salvo et al. 2009; Shaposhnikov et al. 2009 (for Cyg\th X-2) in which the 
feature was fitted as a disk line. However, in \hbox{Cyg\th X-2}, high quality grating spectra 
indicate highly ionized Fe emission with Doppler broadening consistent with the outer ADC, 
not the inner disk and two lines are clearly resolved $\sim$0.3 keV apart, not a Laor 
broadened line (Schulz et al. 2009).

\section{The transient source XTE\th J1701-462}

\begin{figure}[h!]                                                         
\begin{center}
\includegraphics[width=56mm,height=56mm,angle=0]{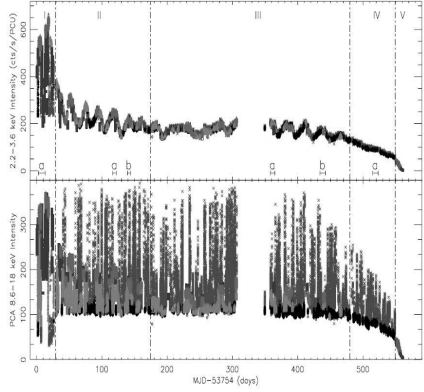}                 
\caption{\footnotesize Lightcurve of XTE\th J1701-462 from Lin et al. (2009).}
\label{}
\end{center}
\end{figure}

The outburst of this source in 2006-7 (Remillard et al. 2006) demonstrated clear
changes in nature, the source moving from Cyg-like to Sco-like and finally to Atoll
like (Lin et al. 2009). It was proposed by Lin et al. that the changes were
due to decreasing $\dot M$. However, this conflicts with the known luminosities of
the classical Z-track sources in which there are no systematic differences in luminosity
between the Cyg-like and Sco-like groups (see Fig. 10).
\begin{figure}[ht!]                                                        
\begin{center}
\includegraphics[width=46mm,height=56mm,angle=270]{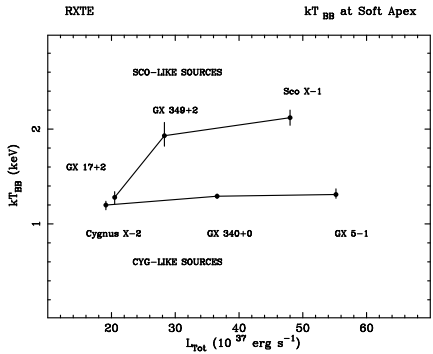}
\caption{\footnotesize
Comparison of the Sco-like sources with the Cyg-like sources emphasizing
a major difference: the  neutron star blackbody temperature $kT$ at the Soft Apex.}
\label{}
\end{center}
\end{figure}

\section{The Sco\th X-1 like sources} 

The main apparent differences displayed by the Sco\th X-1 like sources are seen in Fig. 1.
Spectral fitting using high quality {\it RXTE} data reveals the major physical difference
that at a corresponding point on the Z-track, specifically the Soft Apex,
the neutron star blackbody is very much hotter (Fig. 10).
\begin{figure}[h!]                                                         
\begin{center}
\includegraphics[width=56mm,height=56mm,angle=270]{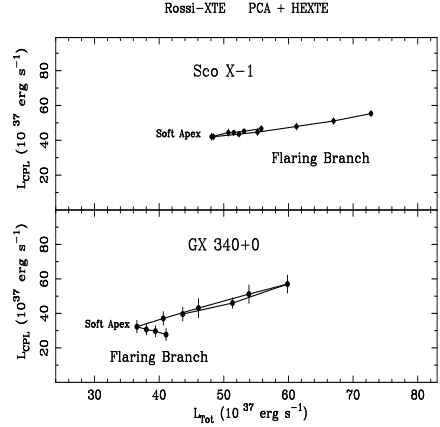}
\caption{\footnotesize
Comparison of flaring in Sco\th X-1 and the Cyg-like source GX\th 340+0.}
\label{}
\end{center}
\end{figure}
\begin{figure}[ht!]
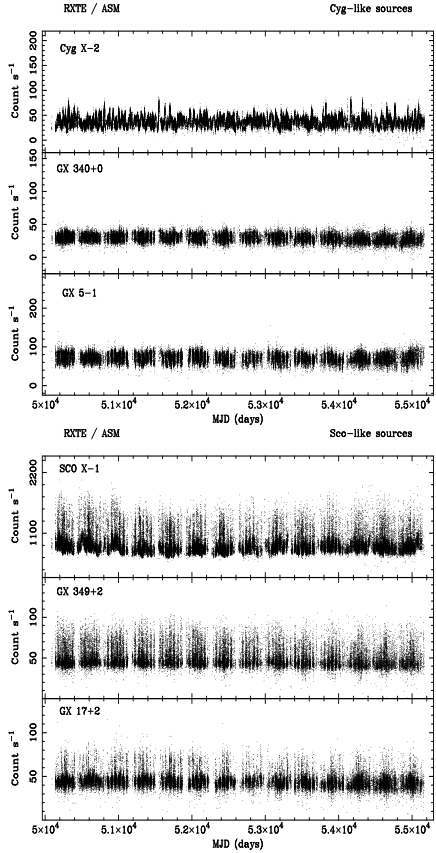
                                                          
\begin{center}
\includegraphics[width=56mm,height=56mm,angle=270]{church_2011_01_fig12a}    
\includegraphics[width=56mm,height=56mm,angle=270]{church_2011_01_fig12b}    
\caption{\footnotesize
The prevalence of flaring in the Sco-like sources (lower panel) compared with the Cyg-like 
sources (upper panel).}
\label{}
\end{center}
\end{figure}

A second difference relates to the nature of flaring. In the Cyg-like sources the constancy
of $\dot M$ was suggested by the constancy of the dominant ADC luminosity component; in
the Sco-like sources $L_{\rm ADC}$ is not constant but increases strongly suggesting that
on both NB and FB $\dot M$ increases.

Our recent work has investigated these effects by the use of the {\it RXTE} ASM over the
full 15 years' life on the satellite. In Fig. 12 we compare the longterm lightcurves
of the Cyg-like and the Sco-like sources. It can be seen that flaring is almost non-stop
and strong in the Sco-like sources while in the Cyg-like sources it is weak and infrequent.
The Sco-like source GX\th 17+2 appears transitional between the two groups (Figs. 10 and 12).

\subsection{Model for the Sco-like sources}

The almost non-stop flaring in the Sco\th X-1 like sources seen in the {\it RXTE} ASM
suggests an explanation of the higher measured neutron star temperature in these sources
of $kT \sim$ 2 keV compared with $kT \sim$ 1 keV in the Cyg-like sources. We propose that
the FB consists of both unstable nuclear burning combined with increase of $\dot M$.
The flaring clearly provides a source of heating of the neutron star and we can 
show that the radiative cooling time is likely to be too long to allow cooling of the 
neutron star. Assuming that a small fraction of the neutron star outer layers is heated, say 1\%.
The higher value of $kT$ compared with the Cyg-like sources gives a stored energy of 
$NkT$ where $N$ = stellar mass/particle mass x 0.01 giving an energy of $3\times 10^{46}$ 
erg. If the star radiates at $L_{\rm BB}$ = $4\times 10^{38}$ erg s$^{-1}$, the cooling
time is 2 years; the 1\% assumed is probably too large, but in any event the
cooling time is longer than the time between flares of a few hours.

\begin{figure}[h!]                                                     
\begin{center}
\includegraphics[width=46mm,height=56mm,angle=270]{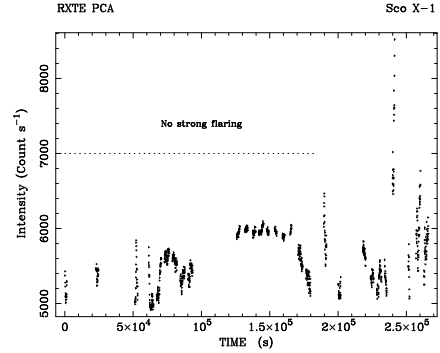}
\caption{\footnotesize Sco\th X-1 without flaring: {\it RXTE} observation of June 1999.}
\label{}
\end{center}
\end{figure}

Testing this hypothesis is difficult given the prevalence of flaring in theses sources.
However, investigation of the complete {\it RXTE} archive on Sco\th X-1 revealed a single observation
in June 1999 (Fig. 13) in which there was a gap of probably 2.5 days without the normal strong flaring (allowing for
data gaps). Analysis showed a small but significant reduction of neutron star temperature $kT$
at the Soft Apex to 1.8 keV supporting the above model (Fig. 14).

\begin{figure}[h!]                                                              
\begin{center}
\includegraphics[width=56mm,height=56mm,angle=270]{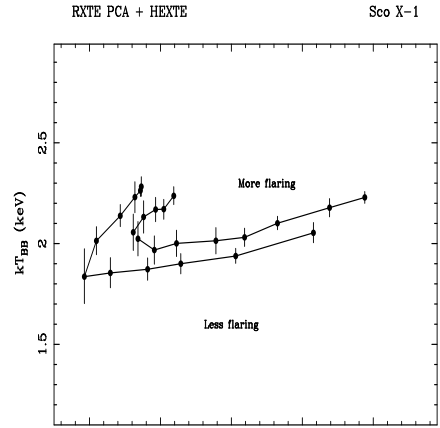}        
\caption{\footnotesize Reduction of neutron star blackbody $kT$ in the observation with reduced flaring.}
\label{}
\end{center}
\end{figure}

We thus propose that the major differences between the Cyg-like and Sco-like sub-groups 
are due to the strong flaring in the latter group. Moreover, spectral studies of the Sco\th X-1 like
group reveal that in flaring, there is both unstable nuclear burning and increase of 
$\dot M$, whereas in the Cyg-like sources flaring consists only of unstable nuclear 
burning. The reason for the almost non-stop flaring is not known, however, it may
reflect corresponding variations in the mass flow to the neutron star.

\section{The Atoll sources}

\begin{figure}[h!]                                                                                      
\begin{center}
\includegraphics[width=56mm,height=56mm,angle=270]{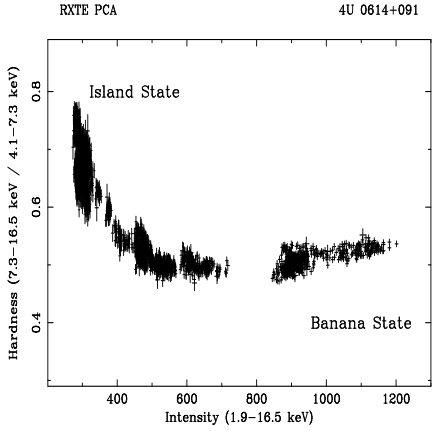}                                
\caption{\footnotesize Movement of a typical Atoll source 4U\th 0614 +091 in hardness-intensity.}       
\label{}
\end{center}
\end{figure}
\begin{figure}[h!]                                                       
\begin{center}
\includegraphics[width=56mm,height=56mm,angle=270]{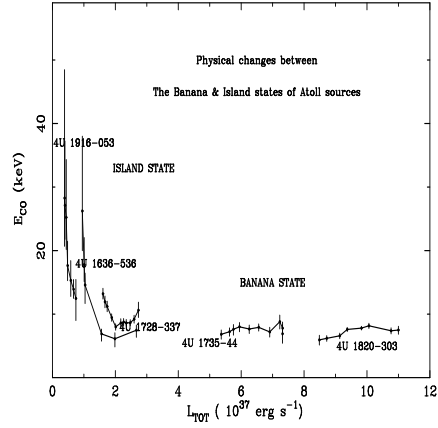}
\caption{\footnotesize
The nature of the Island and Banana States: measured values of the Comptonization high energy cut-off.
}
\label{}
\end{center}
\end{figure}

Finally, the approach of spectral modelling assuming an extended ADC has also
produced an explanation of the Banana and Island states in the lower luminosity Atoll
sources. Fig. 15 shows a typical source in hardness-intensity in which the source moves
from higher luminosity along the Banana state till at the lowest intensity the
hardness increases substantially and the source moves to the Island State.

The results of our recent survey of Atoll sources reveal that the changes in hardness are essentially
due to a single factor: the high energy cut-off $E_{\rm CO}$ of the Comptonized emission (Fig. 16).
In brighter sources, or brighter states of the same source, $E_{\rm CO}$ is low at a few keV; however
at low luminosities $E_{\rm CO}$ can rise by a factor of 10 or more. The ADC electron temperature
$T_{\rm e}$ can be derived from $E_{\rm CO}$ and in Fig. 17, we compare this with the measured neutron star
temperature $kT$. It is clear that in all sources more luminous than $2\times 10^{37}$ erg s$^{-1}$
the ADC is maintained at a low $T_{\rm e}$ equal to $kT$, but for lower luminosities, $E_{\rm CO}$
and so $T_{\rm e}$ become very large due to an unknown coronal heating process.

\subsection{Model for the Atoll sources}. 

We thus propose that the Banana state consists of variations of X-ray intensity
as $\dot M$ changes corresponding to the Normal Branch of the Z-track sources. 
For low $\dot M$ at $L$ $<$ $2\times 10^{37}$ erg s$^{-1}$, the cut-off energy 
rises to 50 keV or more because $T_{\rm e}$ rises leading to the observed strong hardness increase
of the Island State.

\begin{figure}[h!]                                                             
\includegraphics[width=66mm,height=56mm,angle=0]{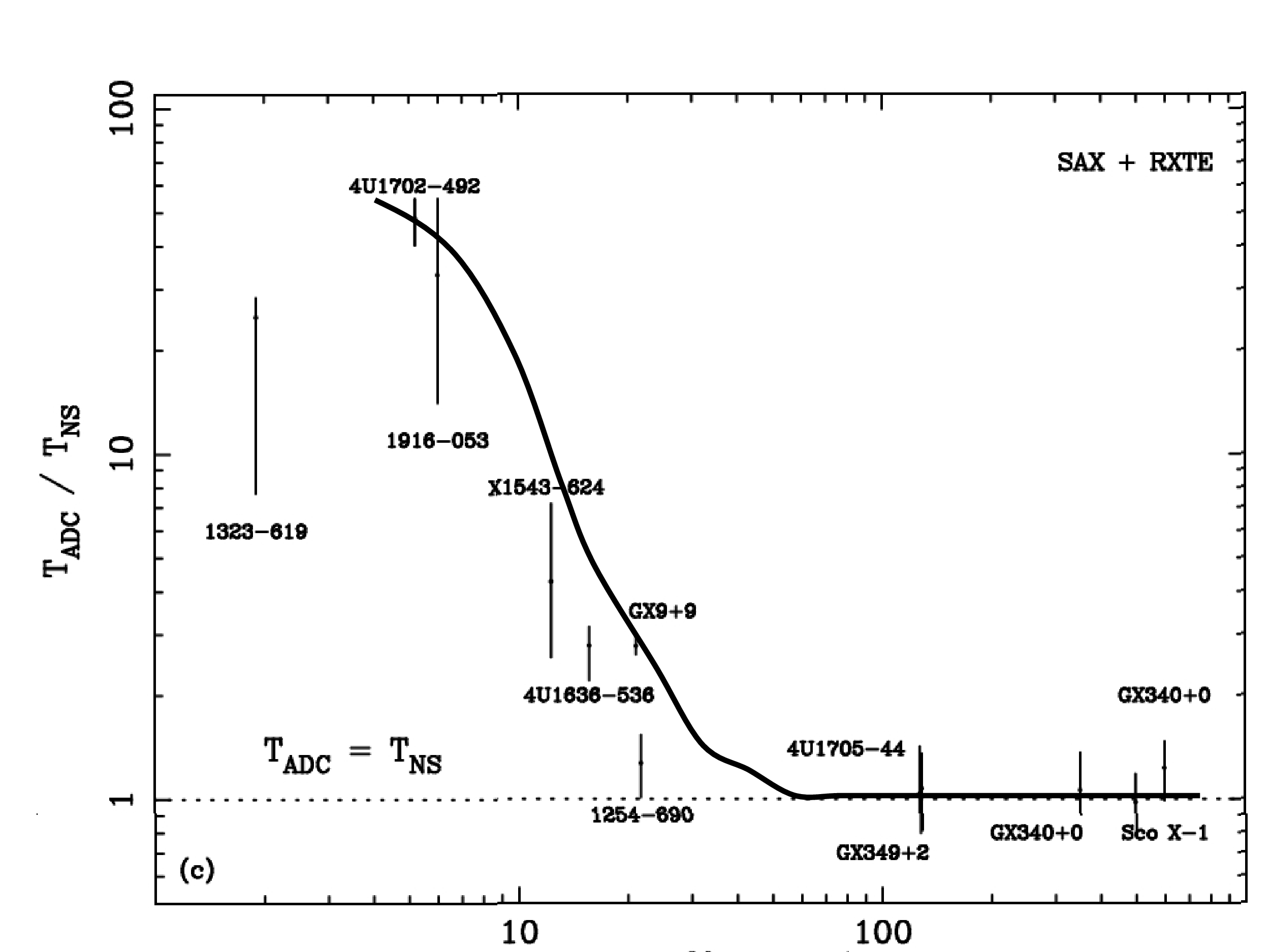}         
\caption{\footnotesize LMXB survey: ratio of ADC electron temperature to neutron star temperature.}
\label{}
\end{figure}

A Flaring Branch is not possible in the Atoll sources as $\dot M$ is many times too small
to permit unstable burning with a critical $\dot m$ equivalent to the high $\dot M$
of these sources. Thus only two tracks, not three, are seen in hardness-intensity.
Of course, in the lower $\dot m $ r\'egime, unstable burning of a different
type: X-ray bursting, is possible.

\section{Conclusions: a Unified LMXB model}

Based on assuming an extended ADC we find that in the Cyg-like Z-track sources
the NB consists of increase of mass accretion rate (soft apex to hard apex) while
flaring is unstable nuclear burning on the neutron star. In the Sco-like sources
a higher neutron star temperature results from almost continuous flaring which
is a combination of $\dot M$ increase and unstable burning. In the Atoll sources
the Banana State has varying $\dot M$ like the Normal Branch, while the Island State
is unique to Atoll sources due to a high cut-off energy resulting from heating in
the corona. The Flaring Branch does not exist but is replaced by X-ray bursting.

\begin{acknowledgements}
This work was supported by the Polish Ministry of
Science and Higher Education grant 3946/B/H03/2008/34.
\end{acknowledgements}

\bibliographystyle{aa}

\vskip 4 mm
\noindent {\bf DISCUSSION}

\vskip 2mm
\noindent {\bf MARAT GILFANOV:} The energy of hydrogen fusion is of the order of
1/30 of the gravitational energy release. Is this consistent with the average 
luminosity of flares ?

\vskip 2mm
\noindent {\bf MICHAEL CHURCH:} In the Cyg-lke sources the flare energy is consistent with 
unstable burning of matter accumulating on the neutron star. In the Sco-like sources it 
appears that flaring is a combination of 
$\dot M$ increase and unstable burning so this does not have to supply 
all of the increased luminosity.


\vskip 2 mm
\noindent {\bf THOMAS BOLLER:} You said that the unstable burning is related to an
increased accretion rate. Are the processes physically connected ?

\vskip 2 mm
\noindent {\bf MICHAEL CHURCH:} 
It is not clear what causes non-stop flaring in the Sco-like sources, or whether
burning is triggered by accretion variations or by other changes such as in
emitting area on the neutron star.

\end{document}